\documentclass[twoside,twocolumn,9pt]{article}
\usepackage{extsizes}
\usepackage[super,sort&compress,comma]{natbib} 
\usepackage[version=3]{mhchem}
\usepackage[left=1.5cm, right=1.5cm, top=1.785cm, bottom=2.0cm]{geometry}
\usepackage{balance}
\usepackage{times,mathptmx}
\usepackage{amsmath,amsfonts,amssymb}
\usepackage{sectsty}
\usepackage{graphicx} 
\usepackage{lastpage}
\usepackage[format=plain,justification=justified,singlelinecheck=false,font={stretch=1.125,small,sf},labelfont=bf,labelsep=space]{caption}
\usepackage{float}
\usepackage{fancyhdr}
\usepackage{fnpos}
\usepackage[english]{babel}
\addto{\captionsenglish}{%
	
}
\usepackage{textgreek}
\usepackage{array}
\usepackage{droidsans}
\usepackage{charter}
\usepackage[T1]{fontenc}
\usepackage[usenames,dvipsnames]{xcolor}
\usepackage{setspace}
\usepackage[compact]{titlesec}
\usepackage{xcolor}
\usepackage{textcomp}

\usepackage{xr}
\definecolor{cream}{RGB}{222,217,201}
\newcommand{\Ic}{\ensuremath{I_{\rm c}}}

\newcommand{\br}{\ensuremath{\mathbf{r}}}
\newcommand{\bs}{\ensuremath{\mathbf{s}}}

\newcommand{\degree}{\ensuremath{^\circ}}

\newcommand{\Eq}[1]{Eq.\,\ref{#1}}
\newcommand{\Fig}[1]{Fig.\,\ref{#1}}
\newcommand{\Sec}[1]{Sec.\,\ref{#1}}
\newcommand{\Tab}[1]{Table \,\ref{#1}}
\newcommand{\Onlinecite}[1]{Ref.\,\hspace{-1 ex} \nocite{#1}\citenum{#1}} 
\newcommand{\be}{\begin{equation}}
	\newcommand{\ee}{\end{equation}}
\newcommand{\bea}{\begin{eqnarray}}
	\newcommand{\eea}{\end{eqnarray}}

\newcommand{\Ai}{\ensuremath{A_{\rm i}}}
\newcommand{\Ab}{\ensuremath{A_{\rm b}}}
\newcommand{\phb}{\ensuremath{\phi_{\rm b}}}
\newcommand{\ri}{\ensuremath{r_{\rm i}}}

\newcommand{\Iout}{\ensuremath{I_{\rm t}}}
\newcommand{\Iex}{\ensuremath{I_{\rm e}}}

\newcommand{\bk}{\ensuremath{\mathbf{k}}}

\newcommand{\Aph}{\ensuremath{A_\phi}}
\newcommand{\Aphme}{\ensuremath{A_\phi^{\rm m}}}

\newcommand{\np}{\ensuremath{n_{\rm p}}}
\newcommand{\nm}{\ensuremath{n_{\rm m}}}

\newcommand{\nso}{\ensuremath{n_{\rm so}}}
\newcommand{\nwo}{\ensuremath{n_{\rm wo}}}

\newcommand{\Vp}{\ensuremath{V_{\rm p}}}
\newcommand{\Rp}{\ensuremath{R_{\rm p}}}

\newcommand{\Ne}{\ensuremath{N_{\rm e}}}
\newcommand{\Na}{\ensuremath{N_{\rm a}}}
\newcommand{\sigmac}{\ensuremath{\sigma_{\rm c}}}
\newcommand{\sigmad}{\ensuremath{\sigma_\delta}}
\newcommand{\sigmasn}{\ensuremath{\sigma_{\rm s}}}
\newcommand{\sigmab}{\ensuremath{\sigma_{\rm b}}}
\newcommand{\sigmasnz}{\ensuremath{\sigma_{\rm s0}}}

\newcommand{\sigmaoc}{\ensuremath{\sigma_1^{\rm c}}}
\newcommand{\sigmaoA}{\ensuremath{\sigma_1^{\rm A}}}

\newcommand{\Ime}{\ensuremath{I^{\rm m}_\pm}}

\usepackage{multirow}

\usepackage{xpatch} 
\makeatletter
\xpatchcmd{\@ssect@ltx}{\@xsect}{\protected@edef\@currentlabelname{#8}\@xsect}{}{}
\xpatchcmd{\@sect@ltx}{\@xsect}{\protected@edef\@currentlabelname{#8}\@xsect}{}{}
\makeatother

\begin{document}
	
\pagestyle{fancy}
\thispagestyle{plain}
\fancypagestyle{plain}{
	
	\fancyhead[L]{\hspace{0cm}\vspace{1.5cm}}
	\fancyhead[R]{\hspace{0cm}\vspace{1.7cm}}
	\renewcommand{\headrulewidth}{0pt}
}

\makeFNbottom
\makeatletter
\renewcommand\LARGE{\@setfontsize\LARGE{15pt}{17}}
\renewcommand\Large{\@setfontsize\Large{12pt}{14}}
\renewcommand\large{\@setfontsize\large{10pt}{12}}
\renewcommand\footnotesize{\@setfontsize\footnotesize{7pt}{10}}
\makeatother

\renewcommand{\thefootnote}{\fnsymbol{footnote}}
\renewcommand\footnoterule{\vspace*{1pt}%
	\color{cream}\hrule width 3.5in height 0.4pt \color{black}\vspace*{5pt}} 
\setcounter{secnumdepth}{5}

\makeatletter 
\renewcommand\@biblabel[1]{#1}            
\renewcommand\@makefntext[1]%
{\noindent\makebox[0pt][r]{\@thefnmark\,}#1}
\makeatother 
\renewcommand{\figurename}{\small{Fig.}~}
\sectionfont{\sffamily\Large}
\subsectionfont{\normalsize}
\subsubsectionfont{\bf}
\setstretch{1.125} 
\setlength{\skip\footins}{0.8cm}
\setlength{\footnotesep}{0.25cm}
\setlength{\jot}{10pt}
\titlespacing*{\section}{0pt}{4pt}{4pt}
\titlespacing*{\subsection}{0pt}{15pt}{1pt}

\fancyfoot{}
\fancyfoot[LO,RE]{\vspace{-7.1pt}}
\fancyfoot[CO]{\vspace{-7.1pt}\hspace{13.2cm}}
\fancyfoot[CE]{\vspace{-7.2pt}\hspace{-14.2cm}}
\fancyfoot[RO]{\footnotesize{\sffamily{1--\pageref{LastPage} ~\textbar  \hspace{2pt}\thepage}}}
\fancyfoot[LE]{\footnotesize{\sffamily{\thepage~\textbar\hspace{3.45cm} 1--\pageref{LastPage}}}}
\fancyhead{}
\renewcommand{\headrulewidth}{0pt} 
\renewcommand{\footrulewidth}{0pt}
\setlength{\arrayrulewidth}{1pt}
\setlength{\columnsep}{6.5mm}
\setlength\bibsep{1pt}

\makeatletter 
\newlength{\figrulesep} 
\setlength{\figrulesep}{0.5\textfloatsep} 

\newcommand{\topfigrule}{\vspace*{-1pt}%
	\noindent{\color{cream}\rule[-\figrulesep]{\columnwidth}{1.5pt}} }

\newcommand{\botfigrule}{\vspace*{-2pt}%
	\noindent{\color{cream}\rule[\figrulesep]{\columnwidth}{1.5pt}} }

\newcommand{\dblfigrule}{\vspace*{-1pt}%
	\noindent{\color{cream}\rule[-\figrulesep]{\textwidth}{1.5pt}} }

\makeatother

\twocolumn[
\begin{@twocolumnfalse}
	\sffamily
	\begin{tabular}{p{18cm} }
		
		  \noindent\LARGE{\textbf{Sizing individual dielectric nanoparticles with quantitative differential interference contrast microscopy}} \\
		\vspace{0.3cm} \\
		
		 \noindent\large{Samuel Hamilton\textit{$^{a}$}, David Regan\textit{$^{a}$}, Lukas Payne\textit{$^{a,b}$}, Wolfgang Langbein\textit{$^{\ast,b}$}, Paola Borri\textit{$^{a}$}} \\		
		\vspace{0.3cm} \\
		 \noindent\normalsize{
		We report a method to measure the size of single dielectric nanoparticles with high accuracy and precision using quantitative differential interference contrast (DIC) microscopy. Dielectric nanoparticles are detected optically by the conversion of the optical phase change into an intensity change using DIC. Phase images of individual nanoparticles were retrieved from DIC by Wiener filtering, and a quantitative methodology to extract nanoparticle sizes was developed. Using polystyrene beads of 100\,nm radius as size standard, we show that the method determines this radius within a few nm accuracy. The smallest detectable polystyrene bead is limited by background and shot-noise, which depend on acquisition and analysis parameters, including the objective numerical aperture, the DIC phase offset, and the refractive index contrast between particles and their surrounding. A sensitivity limit potentially reaching down to 1.8\,nm radius was inferred. As application example, individual nanodiamonds with nominal sizes below 50\,nm were measured, and were found to have a nearly exponential size distribution with 28\,nm mean value. Considering the importance of dielectric nanoparticles in many fields, from naturally occurring virions to polluting nanoplastics, the proposed method could offer a powerful quantitative tool for nanoparticle analysis, combining accuracy, sensitivity and high-throughput with widely available and easy-to-use DIC microscopy.} 
		
	\end{tabular}
	
\end{@twocolumnfalse} \vspace{0.6cm}

]

\renewcommand*\rmdefault{bch}\normalfont\upshape
\rmfamily
\section*{}
\vspace{-1cm}


\footnotetext{\textit{$^{a}$~School of Biosciences, Cardiff University, Cardiff, UK}}
\footnotetext{\textit{$^{b}$~School of Physics and Astronomy, Cardiff University, Cardiff, UK}}

\footnotetext{\textit{${\ast}$~E-mail: langbeinww@cardiff.ac.uk}}





\sloppy

\section{Introduction}
Dielectric nanoparticles (NPs) exist in a multitude of forms and are ubiquitous in our world. They can be naturally occurring (e.g. virions and exosomes), synthetically fabricated (e.g. silica beads, nanodiamonds), or by-products of material degradation (e.g. nanoplastics). These NPs are widely utilised in research and industry, with applications ranging from drug delivery in biomedicine\,\cite{SunACIE14, LiuSmall19} to the fabrication of advanced functional materials\,\cite{DangAM13}. A key requirement for all these applications is the knowledge of the NP size. For example, in cell biology it is well known that the uptake of a NP by the plasma membrane, and the subsequent intracellular trafficking route, tightly depends on the NP size, which in turn is a crucial parameter in the use of NPs as vehicles for drug delivery and therapeutics\,\cite{SunACIE14, LiuSmall19}.

Different from metallic particles, dielectric NPs are typically not electron dense, hence their sizes are more challenging to measure with electron microscopy (EM), the industry-standard technique for NP characterisation. EM analysis is also expensive and typically "low-throughput" since only a limited number of NPs can be examined in one field of view under vacuum (for a review on NP characterisation methods see \Onlinecite{MourdikoudisNanoscale18}). To that end, the use of wide-field optical microscopy to determine the size of individual NPs offers many advantages, including simplicity, low cost, high speed and high throughput, with hundreds of individual NPs rapidly imaged in one field of view, under ambient conditions. However, the spatial resolution of an optical microscope is limited by light diffraction (usually to about 200\,nm), typically larger than the size of the investigated NPs. In other words, differently from EM, optical microscopy methods cannot directly resolve NP dimensions. On the other hand, they can exploit the physical relationship between measurable optical properties and NP sizes to accurately determine the latter. Using this concept, we recently showed that NP sizes could be determined from the optical absorption ($\sigma_{\rm abs}$) and scattering cross-section ($\sigma_{\rm sca}$) of individual NPs measured by wide-field extinction microscopy, with uncertainties down to about 1\,nm in diameter\cite{PayneNS20}. 

While optical extinction microscopy is in principle applicable to any NP material, for particle sizes much smaller than the light wavelength (dipole limit) the technique is practically useful only when NPs exhibit significant optical absorption, such as gold\cite{PayneNS20} and silver NPs\cite{WangNSA20}. This is because $\sigma_{\rm abs}$ scales with the NP volume while $\sigma_{\rm sca}$ scales with the square of the NP volume, severely penalising small NPs which are not absorptive. In other words, by measuring the magnitude of $\sigma_{\rm ext}=\sigma_{\rm abs}+\sigma_{\rm sca}$ one can be sensitive to small NP sizes only if particles are strongly absorbing such that $\sigma_{\rm ext}\sim\sigma_{\rm abs}$. For example, using gold NPs, we have demonstrated a sensitivity limit down to sizes of 2\,nm diameter\cite{PaynePRAP18}.   

Dielectric NPs exhibit a negligible absorption in the visible wavelength range and are weakly scattering. Hence, for these NPs, optical sizing by extinction microscopy is less suited and a different optical contrast method is required. Notably, it is possible to achieve an image contrast proportional to the NP volume using interferometric approaches. For example, it has been recently shown that weakly scattering single dielectric nanoparticles (including biological macromolecules) can be detected with high sensitivity by means of interferometric scattering microscopy (iSCAT) \cite{YoungS18,TaylorNL19}. 

One of the simplest interferometric techniques to generate an image contrast scaling with the NP volume exploits the conversion of the optical phase change of transmitted light introduced by the sample into an amplitude change, by means of differential interference contrast microscopy\,\cite{NomarskiJPRP55}. Briefly, DIC uses a Nomarski prism to split the two linear light polarization components in direction in the condenser back focal plane (BFP), creating a shear distance in the image plane. The two components are recombined by a second prism in the objective BFP. By choosing a shear distance comparable to the spatial resolution of the system, an intensity changing linearly with the differential of the transmitted optical phase along the shear direction is created. Since the phase is proportional to the thickness of an object, the intensity is proportional to the thickness slope. This is akin to the brightness of a modulated surface height under oblique illumination, and thus provides an intuitively interpretable image. The differential contrast also provides optical sectioning. Notably, DIC is widely available in most commercial optical microscopes and is commonly used.

To create quantitative phase information from DIC, various methods have been developed in terms of image acquisition and analysis, eventually resulting in a spatially-resolved map of the optical phase, integrated from the differential phase. For example, the acquisition of images for two orthogonal shear directions and four phase offsets, with subsequent Fourier-space phase integration, was simulated by Arnison {\it et al.}\,\cite{ArnisonJM04} and later experimentally demonstrated by King {\it et al.}\cite{KingJBO08} and Duncan {\it et al.}\cite{DuncanJOSAA11}. Alternatively, using only a focussed and a defocussed DIC image, phase retrieval was shown via the transport of intensity equation\,\cite{KouACP10}. To simplify the acquisition of two shear directions without sample rotation, two orthogonal Nomarski prisms and polarization control can be employed\,\cite{ShribakJBO17}, and axially-offset circularly-polarized DIC was shown \cite{DingOE19}. Moreover, using only a single shear direction, Wiener filtering was demonstrated to be effective in extracting phase images\,\cite{MunsterJM97}, and iterative phase reconstruction \cite{KoosSR16} can further improve the results. By exploiting quantitative DIC (qDIC) with Wiener filtering, we have previously shown that the thickness of lipid bilayers could be measured with a precision of 0.1\,nm \cite{ReganL19,ReganSPIE19}. Furthermore, by directly fitting the DIC contrast without phase integration, the lamellarity of giant lipid vesicles was quantified \cite{McPheeBJ13}. 

The use of qDIC to measure the volume of individual dielectric NPs was proposed by us in an earlier work on single nanodiamonds\,\cite{PopeNNa14}. However, there the investigated nanoparticles were rather large (200-500\,nm diameter) hence the challenge to measure small dielectric NPs with this method was not addressed, neither the detection sensitivity limit, nor the precision or accuracy of the technique was quantified. In this work, we have characterised the application of qDIC for sizing single dielectric NPs and determined the precision and accuracy of the method, depending on the acquisition parameters. As an application example, we show sizing of individual nanodiamonds with only 28\,nm mean size.

\section{Methods}
\subsection{Samples}
\label{sec:Samples}
For calibration of the qDIC method, polystyrene (PS) beads, having a nominal radius of 100\,nm with 3\% standard deviation, were purchased (Alpha Nanotech Colloidal PS Beads NP-PA07CPSX78). These PS beads were dispersed in water and drop cast onto ($24\times24$)\,mm$^2$ \#1.5 coverslips (Menzel Gl\"aser). 
After drop casting and drying, the beads were immersed in oil by pipetting 20\,\textmu l onto the coverslip. To avoid the formation of air bubbles, the samples were then degassed in a vacuum for 10 minutes immediately before adding a microscope slide and sealing the borders using clear nail varnish.  Two types of oil were used, namely water immersion oil (Zeiss, Immersol W 2010) of refractive index $\nwo=1.334$, and silicone oil (Sigma Aldrich, AP 150 Wacker) of index $\nso=1.518$. Prior to use, all glass slides and coverslips were cleaned as follows. First, coverslips and slides were immersed in toluene and sonicated for 20 minutes, followed by being immersed in acetone and sonicated for 20 minutes. Next, they were immersed in deionized (DI) water which was then boiled for 3 minutes. Finally, slides and coverslips were immersed in a 30\% hydrogen peroxide solution, and again sonicated for 20 minutes. After cleaning, slides and coverslips were kept in a refrigerator in the hydrogen peroxide solution, until needed. Nanodiamonds (NDs) were purchased from Microdiamant with nominal sizes $(0-50)$\,nm (MSY $0-0.05$ micron), $(0-150)$\,nm, (MSY $0-0.15$ micron), and $(0-250)$\,nm (MSY $(0-0.25)$ micron). Purchased NDs were purified in-house to remove sp$^2$ graphitic bonds from the surface, by immersion in sulfuric acid for 2 hours, followed by air annealing at 600$^\circ$C for 5 hours. Nanodiamonds deposited onto glass were prepared in the same way as described above for PS beads, using silicone oil as surrounding medium. 

\subsection{Optical Setup}
\label{sec:Setup}

DIC images were obtained using an inverted Nikon Ti-U microscope. Samples were illuminated using a 100\,W halogen lamp (Nikon V2-A LL 100\,W) followed by a Schott BG40 filter to remove wavelengths above 650\,nm (for which the DIC polarisers are not suited) and a Nikon green interference filter (Nikon GIF), to define the wavelength range centred at 550\,nm and having a full-width at half maximum (FWHM) of 53\,nm. This illumination was then passed through a de-S\'enarmont compensator (a rotatable linear polariser and quarter-wave plate, Nikon T-P2 DIC Polariser HT MEN51941) and a Nomarski prism  (Nikon N2 DIC module MEH52400 or MEH52500) and focused onto the sample by a condenser of 0.72\, numerical aperture (NA) or 1.34\,NA (part number MEL56100 or MEL41410, respectively). The shear of the Nikon N2 DIC was measured to be $(238 \pm 10)$\,nm. The objectives used were a $20\times$, 0.75\,NA planapochromat (MRD00205) in conjunction with the 0.72 NA condenser and a $1.5\times$ tube lens, and a $60\times$ 1.27\,NA water immersion planapochromat (MRD70650) or $100\times$ 1.45\,NA planapochromat (MRD01905), in conjunction with the 1.34\,NA condenser and a $1\times$ tube lens. After the objectives, light passes through a suited Nomarski prism (DIC sliders MBH76220, MBH76264, and MBH76190, respectively) and a linear polariser (Nikon Ti-A-E DIC analyser block MEN 51980). Images were detected by a Hamamatsu Orca 285 CCD camera (18,000 electrons full well capacity, 7 electrons read noise, and 4.6 electrons per count, 12 bit digitizer, $1344 \times 1024$ pixels, pixel size 6.45\,$\mu$m, 192 counts offset).

The NA of the condenser lens was matched to that of the chosen objective, with the maximum NA of 1.34 used for the 1.45\,NA objective, and the maximum of 0.72 for the 0.75\,NA objective. The microscope was adjusted for K\"ohler illumination, and the field aperture was set to be slightly larger than the imaged sample region.

A sequence of \Na\ frames (up to 256) were acquired, with 120\,ms exposure time per frame (given by highest stably achievable frame rate), with the lamp intensity adjusted to provide a mean intensity of about 3000 counts (13000 photoelectrons) per pixel.
Data for de-S\'enarmont polarizer angles $\pm\theta$ as well as zero were taken to enable qDIC analysis, for $\theta$ of 15, 30, and 45 degrees. Images with opposite angles were taken in close temporal sequence to minimize drift between both data. The rotation of the de-S\'enarmont polarizer was motorized (by a home-built modification) to improve positioning speed and reproducibility. 

\subsection{qDIC analysis} \label{sec:qDICana}
In order to obtain quantitative phase information, we follow the analysis described in \Onlinecite{ReganL19,PopeNNa14} briefly summarised here for clarity. The transmitted intensity image in DIC can be expressed as
\be\label{eq:Iout} \Iout(\br,\psi)=\frac{\Iex}{2} \left[1-\cos\left(
\psi -\delta(\br)\right)\right], \ee
with the excitation intensity \Iex, the position in the sample plane
$\br$, the phase offset  $\psi$, and the difference $\delta(\br)$ of the
optical phase shift $\phi$ for the two beams that pass through
the sample in two adjacent points separated by the shear vector
$\bs$. This is expressed as
\be\label{eq:delta} \delta(\br)=\phi(\br+\bs/2)- \phi(\br-\bs/2).\ee
To reduce the influence of a residual spatial dependence of \Iex, which includes inhomogeneities in illumination and detection, we acquire two images at opposite angles $\theta$ of the de-S\'{e}narmont polarizer, providing the 
intensities $I_\pm=\Iout(\br,\pm \psi)$ with
$\psi=2\theta$. The contrast image is then defined as
\be\label{eq:Ic}
\Ic\left(\br\right)=\frac{I_+(\br)-I_-(\br)}{I_+(\br)+I_-(\br)}.
\ee
By combining \Eq{eq:Iout} and \Eq{eq:Ic}, we obtain
\be\label{eq:contrasti}
\Ic\left(\br\right)=\frac{\sin(\psi)\sin(\delta)}{\cos(\psi)\cos(\delta)-1}
\ee
which, for $0\leq\psi\pm\delta\leq\pi$, can be solved analytically\cite{PopeNNa14},  yielding 
\be\label{eq:exact} \delta(\br) = \arcsin\left(\Ic\frac{\cos(\psi)\sqrt{1-\Ic^2}-1}{\sin(\psi)(1+\Ic^2\cot^2(\psi))}\right)\,.
\ee
To extract the phase $\phi$ from $\delta$, a Wiener deconvolution in the Fourier domain of wavevector \bk\ is used. \Eq{eq:delta} is written in the Fourier domain as
\be \mathcal{F}(\delta(\br)) = \xi(\bk)\mathcal{F}(\phi(\br))\,,\ee
with $\xi(\bk) = 2i \sin (\bk\cdot\bs/2)$ and $\mathcal{F}$ denoting the Fourier transform. Using Wiener deconvolution with a signal to noise parameter $\kappa$, we retrieve the phase using 
\be \label{eq:Wiener} \phi(\br) \approx \mathcal{F}^{-1}\left(  \frac{\mathcal{F}(\delta(\br))}{\xi(\bk)+(\kappa\xi(\bk)^*)^{-1}}\right)\,,\ee
where the $^*$ denotes the complex conjugation. 

To analyze particle volumes, the phase $\phi(\br)$ is then spatially integrated over a circular area centred at the NP (in a similar way as we introduced for extinction images \cite{PayneAPL13, PayneFD15}) using a dual radius method, as follows. Firstly, particle positions are determined by maxima of $\phi$.  The background phase for a given particle is determined as the phase over the area \Ab\ within the distance \ri\ and $2\ri$ from the particle position, namely
\be \label{eq:phb} \phb=\Ab^{-1}\int_{\Ab}\phi(\br) d\br\,. \ee
The measured integrated phase \Aphme\ over the particle is then calculated over an area \Ai\ with a distance below \ri\ from the particle position, using 
\be \label{eq:Aph} \Aphme =\int_{\Ai}(\phi(\br)-\phb)d\br\ee
where $d\br$ is the area element, $dx dy$ in cartesian coordinates. When considering the optical phase difference created for light of wavelength $\lambda$ by a particle of refractive index \np\ surrounded by a medium of index \nm, we can introduce the particle thickness $t(\br)$, leading to a phase difference to the surrounding of
\be \phi(\br) = \frac{2\pi}{\lambda}(\np - \nm)t(\br).\ee
Evaluating \Eq{eq:Aph} for this phase difference, we find
\be \label{eq:AphV} \Aph=\frac{2\pi}{\lambda}(\np - \nm)\Vp\ee
with the particle volume
\be \label{eq:Vp} \Vp =\int_{\Ai}t(\br)d\br\ee
located completely inside \Ai. Therefore, \Vp\ can be determined from $\Aph$ knowing the refractive index values and the wavelength. For spherical particles, the volume is determined by their radius \Rp, so that we find  
\be \label{eq:AphR} \Rp =\sqrt[3]{\frac{3 \lambda \Aph}{8\pi^2(\np - \nm)}}\,.\ee
Importantly, the measured phase area \Aphme\ is affected by the finite spatial resolution and the finite $\kappa$ in \Eq{eq:Wiener}, and has to be corrected to obtain \Aph, as we detail later. The software and parameters used for the analysis are described in the SI \Sec{S-sec:software}.

\section{Results and Discussion}

\subsection{PS beads}

\begin{figure*}[t]
	\includegraphics*[width=\textwidth]{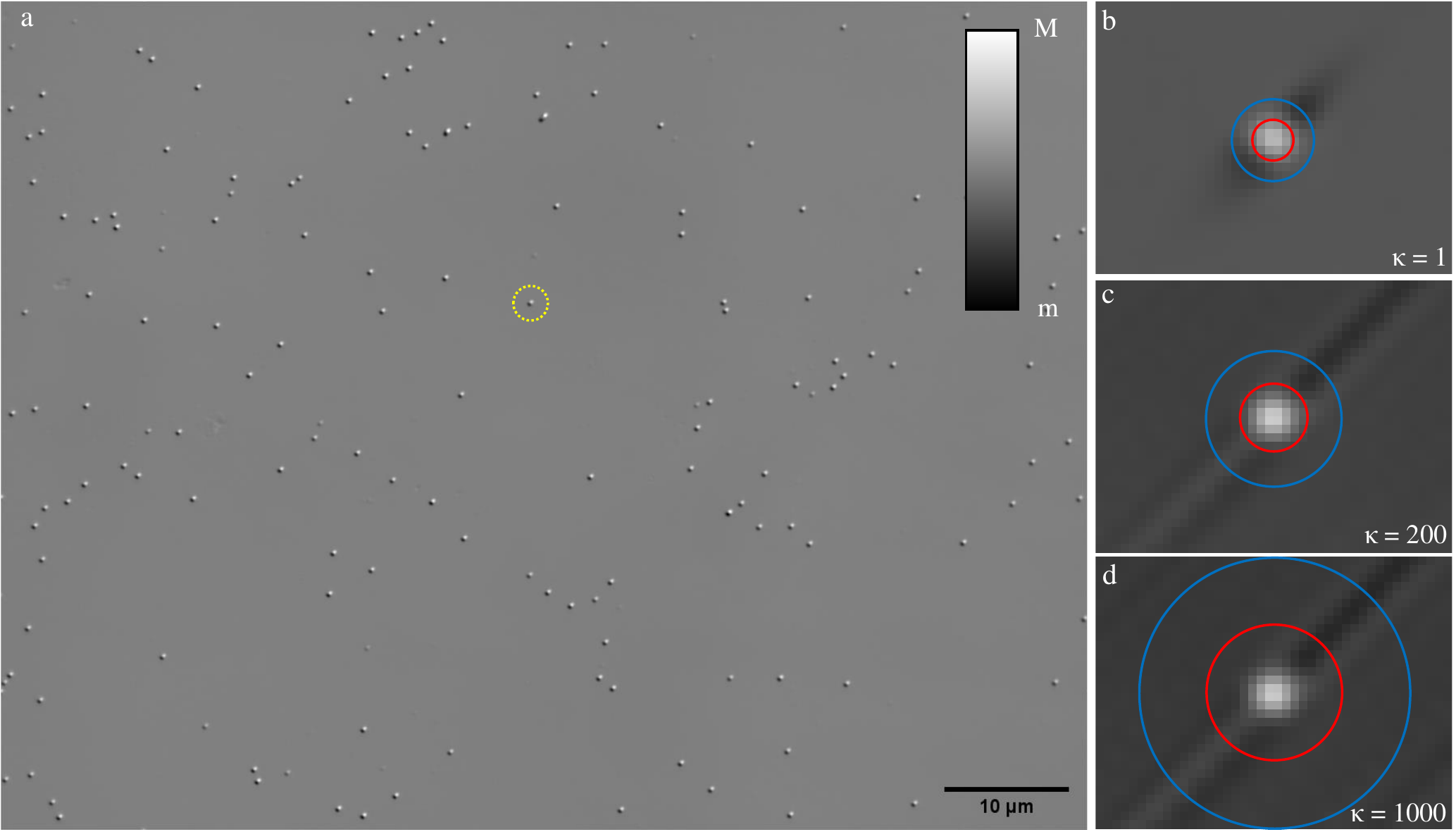}
	\caption{\label{fig:ROI} qDIC microscopy on individual PS beads of nominal 100nm radius, drop cast onto glass and surrounded by silicon oil, imaged with a 1.45\,NA objective and a phase offset of $\psi=30^\circ$. a) $\delta(\br)$ on a grey scale as shown, from  $m = -0.05$ to $M = 0.05$. The shadow cast impression is evident, with the shear $\bs=0.16(1,1)$\,\textmu m in the $(x,y)$ coordinates ($x$ is the horizontal axis and $y$ the vertical in the image). Optical phase maps $\phi(\br)$ showing a region of $(2.71 \times 2.07)$\,\textmu m$^2$ around a selected bead indicated by the dashed circle, for $\kappa=1$  (b, $m = -0.015$ to $M = 0.03$), $\kappa=200$ (c,  $m = -0.02$ to $M = 0.04$), and $\kappa=1000$ (d, $m = -0.03$ to $M = 0.03$). The red and blue circles have the radii \ri\ and $2\ri$, respectively, with $\ri=2.5,4,8$ pixels in b,c,d, respectively, representing different integration areas \Ai\ and \Ab\ used in the analysis for \Eq{eq:phb} and \Eq{eq:Aph}.}
\end{figure*}\par

PS beads of known radius and refractive index were used as reference standard, to test the accuracy of NP sizing by qDIC. A representative differential phase image $\delta(\br)$ for nominally 100\,nm radius PS beads deposited onto glass and embedded in silicon oil is shown in \Fig{fig:ROI}a using a 1.45\,NA microscope objective at a phase offset of $\psi=30^\circ$ (see Methods for details of the sample and optical set-up). The corresponding  images for the 0.75NA and 1.27NA objectives are shown in the Electronic Supplementary Information (ESI)  \Fig{S-fig:ROI_20x} and \Fig{S-fig:ROI_60x}, respectively. The typical shadow-cast appearance of the individual beads is observed. Note the remarkable absence of blemishes or vignetting on a contrast scale of only $\pm5$\%, which is a result of using the DIC contrast \Eq{eq:Ic}, as compared to individual DIC images (for illustration the $I_+$ image corresponding to \Fig{fig:ROI}a shown in the ESI \Fig{S-fig:ROI_100x_Ip}).

\subsubsection{qDIC Optimization and Calibration}
\label{sec:Calibration}

The qDIC analysis discussed in \Sec{sec:qDICana} uses as parameters the signal to noise ratio $\kappa$ in the Wiener deconvolution, and the area radius \ri\ to evaluate the integrals. To choose the parameter values for best precision and accuracy, the dependence of the measured integrated phase \Aphme\ and its noise is evaluated as function of these parameters. For the discussion, let us consider here data taken on PS beads mounted in silicone oil, for the 1.45NA microscope objective, at a phase offset of $\psi=30^\circ$. The data were analysed for $\kappa$ ranging from 0.5 to $10^5$, and \ri\ from 0.5 to 9 pixels. Representative images of the optical phase $\phi(\br)$ for $\kappa = 1$, 200, and 1000 are shown in \Fig{fig:ROI}b-d around a single bead. 

As $\kappa$ is increased, the extension of spatial features along the shear direction is increasing proportional to $\sqrt{\kappa}$. This is the result of the spatial high pass filter along the shear resulting from the Wiener filter of qDIC, \Eq{eq:Wiener}.
Its cut-off wavevector $\bk_c$ is given by the condition $\kappa|\xi(\bk_c)|^2=1$, which for small $|\xi|$ is approximated by $|\bk_c\cdot\bs|\sqrt{\kappa}=1$, so that $|\bk_c|$ is proportional to $1/\sqrt{\kappa}$. 
While allowing for longer range features to be retrieved, increasing $\kappa$ also increases the noise due to the larger amplification of the data by the filter function for small $|\xi|$. 

Notably, the stripes show here a triplet structure, which is attributed to the asymmetry of the point-spread function for linearly polarised light. As the two sheared components have linear polarization along and across the shear, their spatial elongation is oriented also in this way, resulting in accordingly different shapes of the probed regions. For smaller NA, this asymmetry is reduced, and with it the triplet structure, as can be seen in the results for the 0.75\,NA objective in \Fig{S-fig:ROI_20x} . 

The radius \ri\ instead determines the size of the circular areas \Ai\ and \Ab\ over which the integrals of the optical phase are calculated (\Eq{eq:phb} and \Eq{eq:Aph}), as shown by the red and blue circles in \Fig{fig:ROI}b-d. For \ri\ larger than the cut-off of the Wiener filter, \Ai\ contains also regions of inverted (negative) contrast (see dark tails in \Fig{fig:ROI}b-d), reducing the resulting \Aphme. On the other hand, for \ri\ smaller than the spatial resolution, \Ai\ will not contain the full response and again \Aphme\ will be reduced. Furthermore, the areas scale with $\ri^2$, so that the shot-noise in \Aphme\ will scale with \ri, favoring small \ri\ for high signal to noise ratio (SNR).

\begin{figure}
	\centering
	\includegraphics[width=\columnwidth]{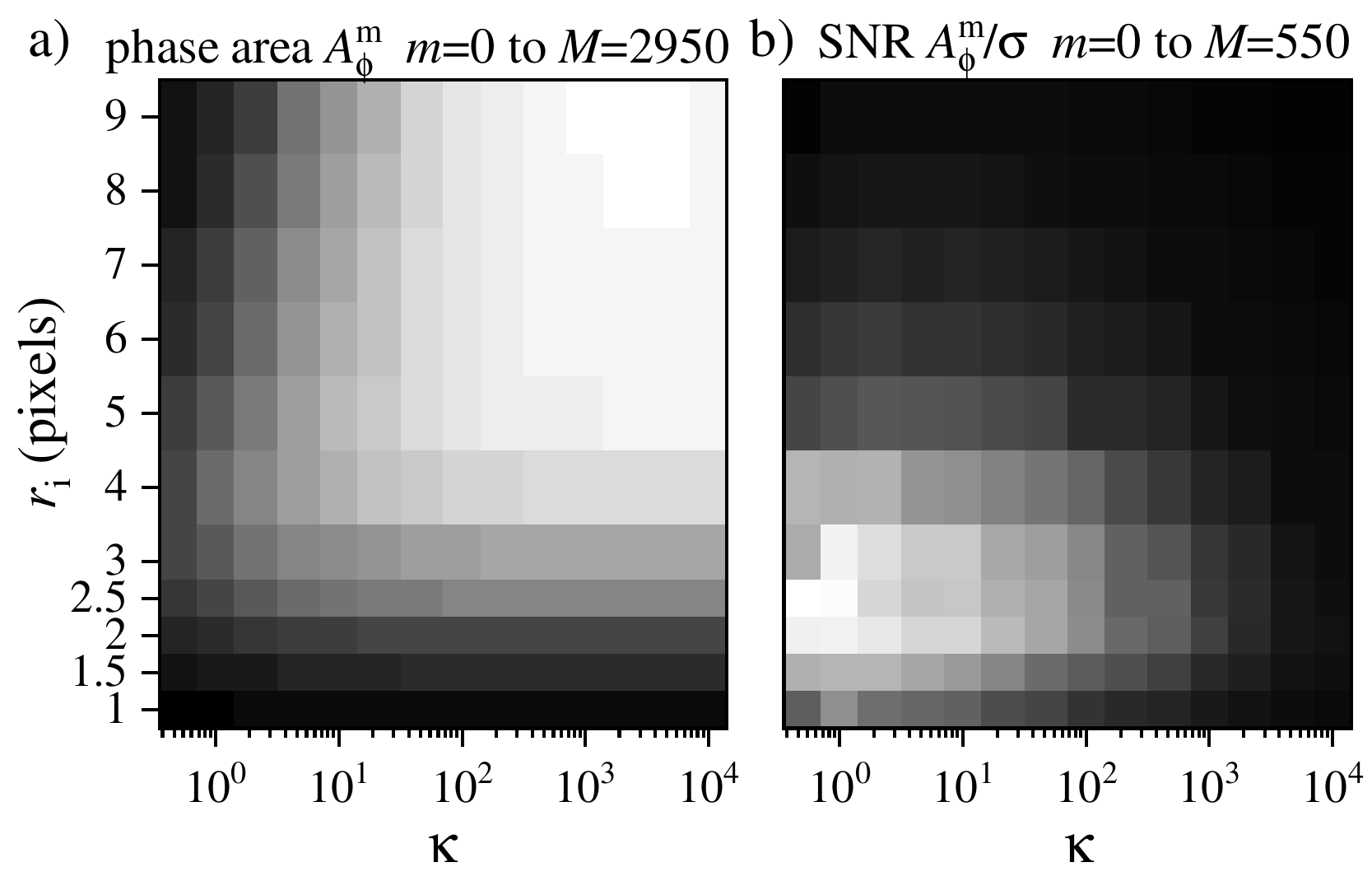}
	\caption{Phase area \Aphme\ (a, $m=0$ to $M=2950$\,nm$^2$) and SNR $\Aphme/\sigma$ (b, from $m=0$ to $M=550$) as function $\kappa$ and \ri\ for a PS bead in silicon oil imaged with the 1.45\,NA objective and a phase offset of $\psi=30^\circ$ as in \Fig{fig:ROI}.}
	\label{fig:kappaRiDep}
\end{figure}

The evaluated \Aphme\ as function of $\kappa$ and \ri\ is given in \Fig{fig:kappaRiDep}a for the bead selected in \Fig{fig:ROI}. We find, in accordance with the above qualitative arguments, that \Aphme\ is increasing steeply with \ri\ up to about 4 pixels, which is the size of the point-spread function (PSF) (see red circle in \Fig{fig:ROI}c). For larger \ri, \Aphme\ reduces for small values of $\kappa$, and increases for large $\kappa$, converging to a stable value for $\kappa>500$ and $\ri>7$. 

We define three $(\kappa,\ri)$ pairs according to the following criteria: the pair that provides the highest SNR (called SN pair), the one for which \Aphme\ converges to the highest value (called C pair), and a compromise choice which still gives a good SNR but a reduced systematic error due to a lower sensitivity to the specific shape of the PSF (called SE). Based on \Fig{fig:kappaRiDep}a, as C pair we use $(\kappa, \ri)=(1000,8)$, where the units of \ri\ are pixels (one pixel has a size of 65\,nm on the sample for these data). To choose the SN pair, we determine the SNR by evaluating \Aphme\ at positions not showing a visible particle in the image, and fit its histogram with a Gaussian to determine its standard deviation $\sigma$, as can be seen in the ESI \Fig{S-fig:BGPhaseArea}. The resulting SNR $\Aphme/\sigma$  corresponding to \Fig{fig:kappaRiDep}a is given in \Fig{fig:kappaRiDep}b.
We find that the SNR is increasing with \ri\ up to about 2 pixels. This can be understood considering that for small \ri, \Aphme\ is scaling with $\ri^2$, while  $\sigma$ scales only with \ri. For larger \ri\ instead, \Aphme\ is saturating or even decreasing, as seen in \Fig{fig:kappaRiDep}a, so that the SNR decreases, due to the increasing $\sigma$. Moreover, for $\kappa$ above 2, for which the Wiener filter cut-off is larger than the PSF, the SNR is decreasing as expected from the  qualitative arguments mentioned previously. The highest SNR is obtained for the SN pair $(1, 2.5)$. 
Finally, for the SE pair we chose a larger \ri\ corresponding to the PSF size, and accordingly the $\kappa$ giving the highest SNR, which is $(200,4)$. This choice reduces systematic errors observed for lower \ri, as will be shown later.  

Note that the values of $\kappa$ and \ri\ for the SN, C and SE pairs depend on the objective and tube lens used, which determine the optical resolution. We report  in the ESI \Aphme\ as function of $\kappa$ and \ri\ for the 0.75 NA and 1.27 NA objectives, see \Fig{S-fig:kappaRiDep20x}a and \Fig{S-fig:kappaRiDep60x}a, with the corresponding SNR $\Aphme/\sigma$, see \Fig{S-fig:kappaRiDep20x}b and \Fig{S-fig:kappaRiDep60x}b, and the resulting parameters for the SN, C and SE pairs.

\subsection{Correction Factors and Polystyrene Bead Radii}
\begin{figure}
	\centering
	\includegraphics[width=\columnwidth]{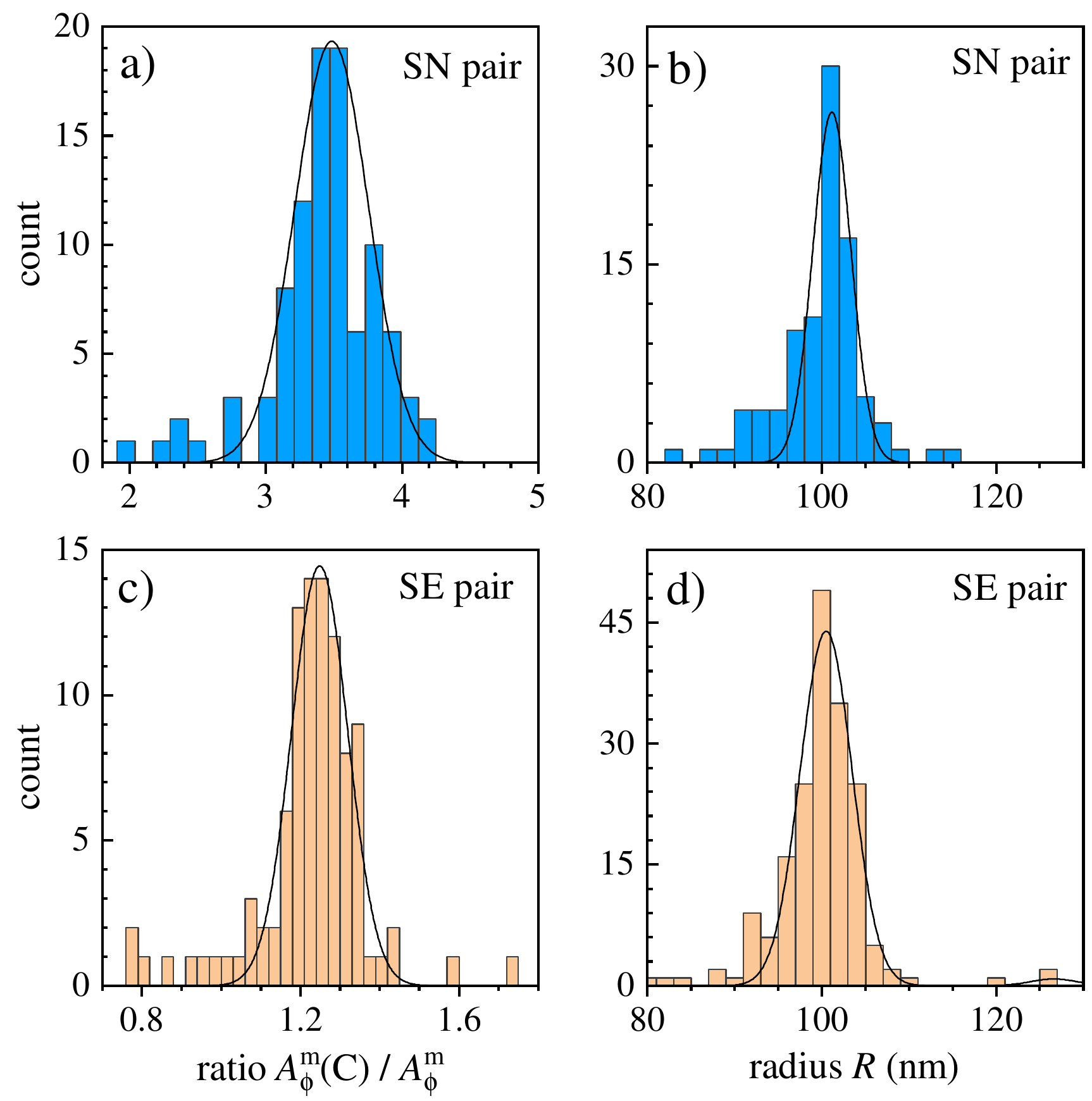}
	\caption{Analysis of PS beads in silicon oil measured using the 1.45\,NA objective and a phase offset of $\psi=30^\circ$ as in \Fig{fig:ROI}. (a,c) Histogram of the ratio of \Aphme\ for the $(\kappa,\ri)$ pair C to \Aphme\ for pair SN (a) or SE (c), with Gaussian fits yielding the mean correction factor $\varrho$. (b,d) Histograms of the resulting bead radii $R$, for the SN (b) and SE (d) pair.}
	\label{fig:CorrFacRadius}
\end{figure}

Since the SN and SE pairs provide a phase area \Aphme\ which is lower than the converged value given by the C pair, we determine correction factors $\varrho$ for these pairs to scale \Aphme\ to the converged value representing \Aph\ (see also \Eq{eq:Aph} and \Eq{eq:AphV}). To do this, each particle's \Aphme\ for the SN and SE pairs was divided by the converged value of the C pair, and the histogram of the resulting ratios was fitted with a Gaussian distribution, to determine center and standard deviation. The distribution obtained for the PS beads mounted in silicone oil imaged using the 1.45\,NA objective is shown in \Fig{fig:CorrFacRadius} for both the SN and SE pair, resulting in correction factors of $\varrho=3.48\pm0.27$ and $\varrho=1.25\pm0.07$, respectively. The correction factors found for all objectives and samples are given in \Tab{tab:ParticleSizes}. Notably, the relative standard deviation of correction factors is generally larger for the SN than the SE pair, showing a reduced systematic error for the SE pair.
The mean correction factor was then used to define the phase area $\Aph=\varrho\Aphme$, which in turn determines the particle volume and corresponding radius from the measurements. Histograms of the resulting radii for the SE and SN pairs are shown in \Fig{fig:CorrFacRadius}b,d. 

To find the mean particle size and standard deviation for each measured radius distribution, the following fit function was used, given by a sum of Gaussian distributions to account for particle aggregates: 
\be	p(R)=\sum_{n=1}^{8}\frac{B_n}{\sigma_n\sqrt{2\pi}}\exp\left(\frac{-(R-R_n)^2}{2\sigma_n^2}\right), \label{eq:FitFunction} \ee
where $R_n=R_1 n^{1/3}$ is the mean radius of an $n$-bead aggregate, and  $\sigma_n=\sigma_{1} n^{-6/2}$ is the standard deviation of this radius, assuming an independent radius variation of the individual beads in the aggregate (a derivation of $R_n$ and $\sigma_n$ is shown in the ESI). Furthermore, for a Poisson distribution of bead numbers in the aggregates, we have 
\be	B_n = B\frac{\lambda^ne^{-\lambda}}{n!}\,,\label{eq:Poisson} \ee
with a normalization $B$ and the average number $\lambda$ of beads per aggregate. Note that $\lambda$ includes the $n=0$ probability, which is not part of the analyzed particles. Fits are shown in \Fig{fig:CorrFacRadius}b and d, yielding the parameters $R_1=101.2$\,nm, $\sigma_1=2.2$\,nm, $\lambda=0.98$ and $B=396.5$ for the SN pair, and $R_1=100.5$\,nm, $\sigma_1=3.0$\,nm, $\lambda=0.03$ and $B=9927$ for the SE pair. Note the quantitative agreement between the measured PS radius and the size provided by the manufacturer within the specified standard deviation. This shows the accuracy of the method, as further discussed later in subsection D. Histograms of the correction factors and bead radii for PS beads in different mounting media and for the various objectives used are shown in the ESI (\Fig{S-fig:CorrFacRadius_20x_WO} to \Fig{S-fig:CorrFacRadius_60x_SO}), with the resulting value of bead radii $R=R_1 \pm \sigma_1$ summarised in \Tab{tab:ParticleSizes}.

\subsection{Background and shot noise}

To characterise the precision of the method, we evaluated the error derived from the noise in the measurements. The noise in the qDIC $\delta$ images in the absence of strong contrast, that is for $\Ic \approx 0$, consists of two components. Firstly, we need to consider the photon shot noise in the measured images $I_\pm$, which depends on the average number of detected photoelectrons per pixel \Ne. For an acquisition consisting of \Na\ frames which are averaged, the shot noise is $\sigmac=1/\sqrt{2\Ne\Na}$ in the DIC contrast \Ic\ (where the factor $1/\sqrt{2}$ accounts for the use of two images in \Ic, see \Eq{eq:Ic}). For typical values used in our work, $\Ne=10^4$ and $\Na=256$, we find $\sigmac=0.04$\%, and for a single frame $\sigmac=0.7$\%. To evaluate the corresponding noise \sigmad\ of $\delta$, which is related to \Ic\ by \Eq{eq:contrasti}, we find
\be \label{eq:sigmad} \sigmac=\sigmad\left|\frac{d\Ic}{d\delta}\right|_{\delta=0}=\sigmad \left| \frac{\sin(\psi)}{1-\cos(\psi)}\right|\,. \ee
We can see that for small offset angles  $0<\psi \ll 1$, \sigmad\ is reduced, by a factor of about $\psi/2$ compared to the noise for $\psi=90\degree$, as discussed previously \cite{McPheeBJ13}. However, smaller $\psi$ also reduces the range which can be retrieved, and the transmitted intensity (\Eq{eq:Iout}) is reduced, requiring longer measurement times or stronger illumination. Furthermore, the non-ideal optical elements used (finite extinction of the polarizers, non-perfect matching of the DIC prisms, birefringence of the objective due to residual strain and oblique transmission) results in deviations of the measured data from the ideal behaviour given by \Eq{eq:Iout}. Most notably, in high quality objectives as used here, light rays incident at large oblique angles, collected and collimated by the objectives, are also at oblique incidence on the lens surfaces of the objectives. The resulting polarization dependent transmission of s and p polarized waves provides a variation of the polarization of the collimated ray after the objective, which depends on its incident direction. As a consequence, the rays are not completely blocked by the polarizer, and a significant transmission at $\psi=0$ can be observed also without sample. We quantify this background transmission as a fraction $\eta$ of \Iex, which was found to be $\eta=0.80$\%, $0.64$\%, and $0.86$\% for the 1.45\,NA, 1.27\,NA, and  0.75\,NA objectives, respectively. Notably, for the smallest $\psi$ used in this work, i.e. $30\degree$, the transmission \Eq{eq:Iout} is only 6.7\% of \Iex, so that the background constitutes a significant fraction (about 6\%) of the ideal transmission without sample. To correct for this residual transmission (i.e. non perfect extinction) in the analysis, we have subtracted this background from the measured intensities \Ime\ to determine $I_\pm$. This equates to using $I_\pm=\Ime-\langle\Ime\rangle 2\eta/(1-\cos(\psi))$ in \Eq{eq:Ic}, where $\langle.\rangle$ denotes the spatial average. 

Other than shot-noise, we have random structures in our samples unrelated to the particles of interest (POI). Notably, the samples that we study consist of a glass coverslip with attached particles, embedded in an immersion oil. We are imaging the glass - immersion oil interface, while other interfaces are out of focus by at least 10\,\textmu m, making them essentially invisible in DIC. Therefore, the background in the absence of POIs originates from unwanted structures at the glass - immersion oil interface. It is thus paramount to use high quality coverslips and clean them properly before attaching the POIs (see sample preparation protocol in Methods section). Even after cleaning, however, there is a remaining surface roughness of a few nanometers which is an intrinsic feature of glass surfaces fabricated by float techniques, due to the thermally excited surface waves at the glass transition during cooling. Since in DIC the contrast at the interface scales with the refractive index difference between glass and immersion oil, optically clearing the interface by matching the refractive indices is an effective way to suppress background from surface roughness. The two immersion oils used in the present work have an index difference to glass of about $0.2$ (water oil) and $<0.002$ (silicon oil). Thus, when using silicon oil, surface roughness is not relevant, while with water oil, the background is dominated by the glass surface roughness.

\begin{figure}
	\includegraphics[width=\columnwidth]{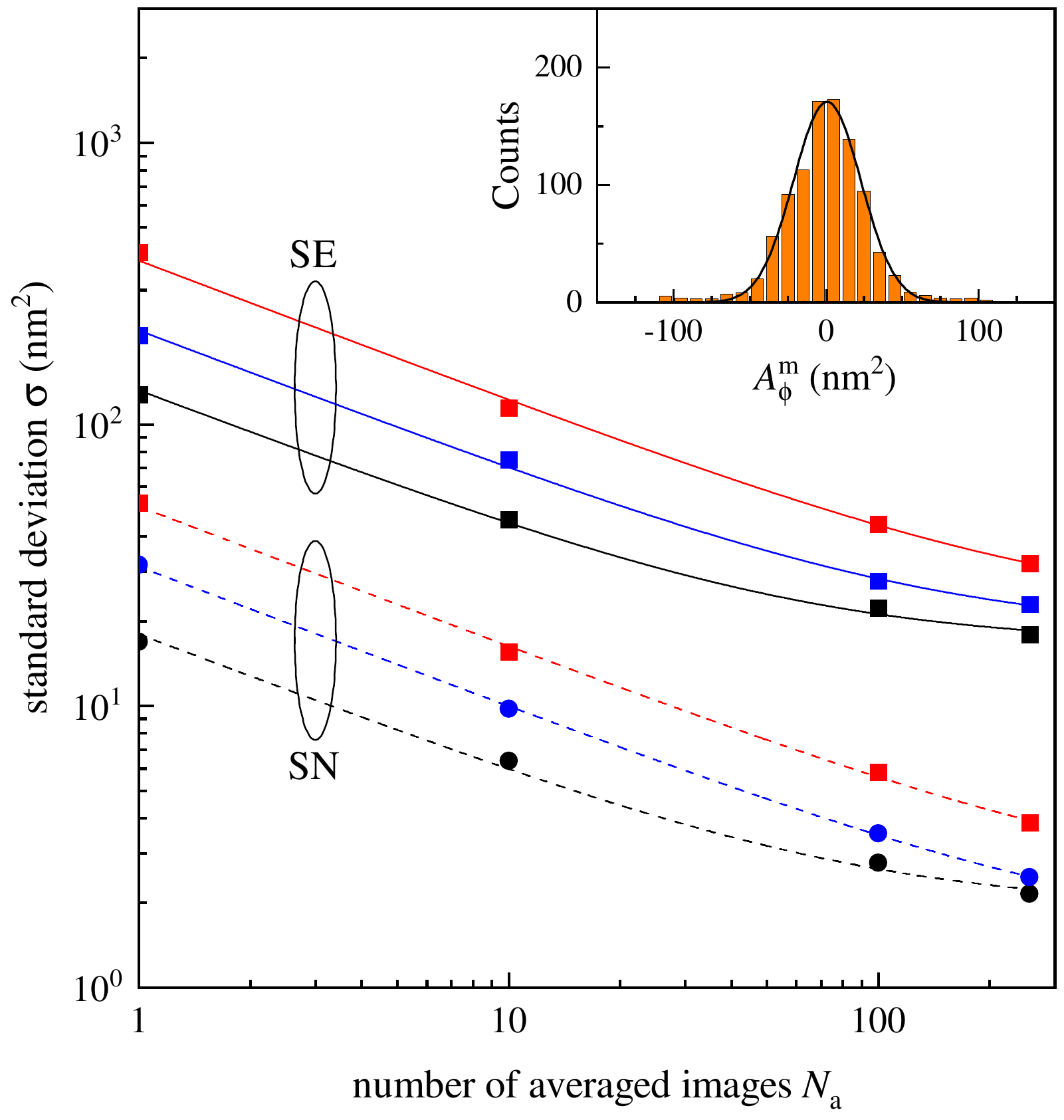}
	\caption{Standard deviation $\sigma$, from the distribution of \Aphme\ in regions of the sample without PS beads, versus number of averages \Na, for PS beads mounted in silicone oil imaged using the 1.45\,NA objective and phase offsets $\psi$ of $30\degree$ (black), $60\degree$ (blue), and $90\degree$ (red). The inset shows the histogram of \Aphme\ for $\Na=100$, analysed using the SE pair imaged at $\psi = 30^\circ$, and the fitted Gaussian distribution (black line).
	}
	\label{fig:BGAverages}
\end{figure}

To unpick the background and shot noise contributions, we determine the noise in \Aphme\ using 1000 points in regions without evident PS beads, which were then analysed with the SN and SE $(\kappa,\ri)$ pairs. A Gaussian function was fitted to the resulting integrated phase area distribution to determine its standard deviation $\sigma$. An example of this histogram is shown in the inset of \Fig{fig:BGAverages}, and the resulting $\sigma$ is shown in \Fig{fig:BGAverages} as function of \Na\ for the 1.45\,NA objective on the sample in silicone oil and phase offsets of $\psi=30,\,60,$ and $90\degree$. We find a decreasing $\sigma$ with increasing \Na, as expected for shot noise, which tends to saturate for $\Na>100$, indicating the background noise limit. We fit this dependence as
\be	\sigma = \sqrt{\frac{\sigmasn^2}{\Na}+\sigmab^2}, \label{eq:shotnoise}\ee
where \sigmasn\ is the shot-noise for a single frame, and \sigmab\ is the background noise due to sample inhomogeneities. 
The resulting \sigmasn\ and \sigmab\ are given in \Tab{tab:ParticleSizes}. 
Similar plots for the other objectives and immersion oils are shown in the ESI \Fig{S-fig:BGAverages_20x_SO} to \Fig{S-fig:BGAverages_100x_WO}. 

\begin{figure}
	\centering
	\includegraphics[width=\columnwidth]{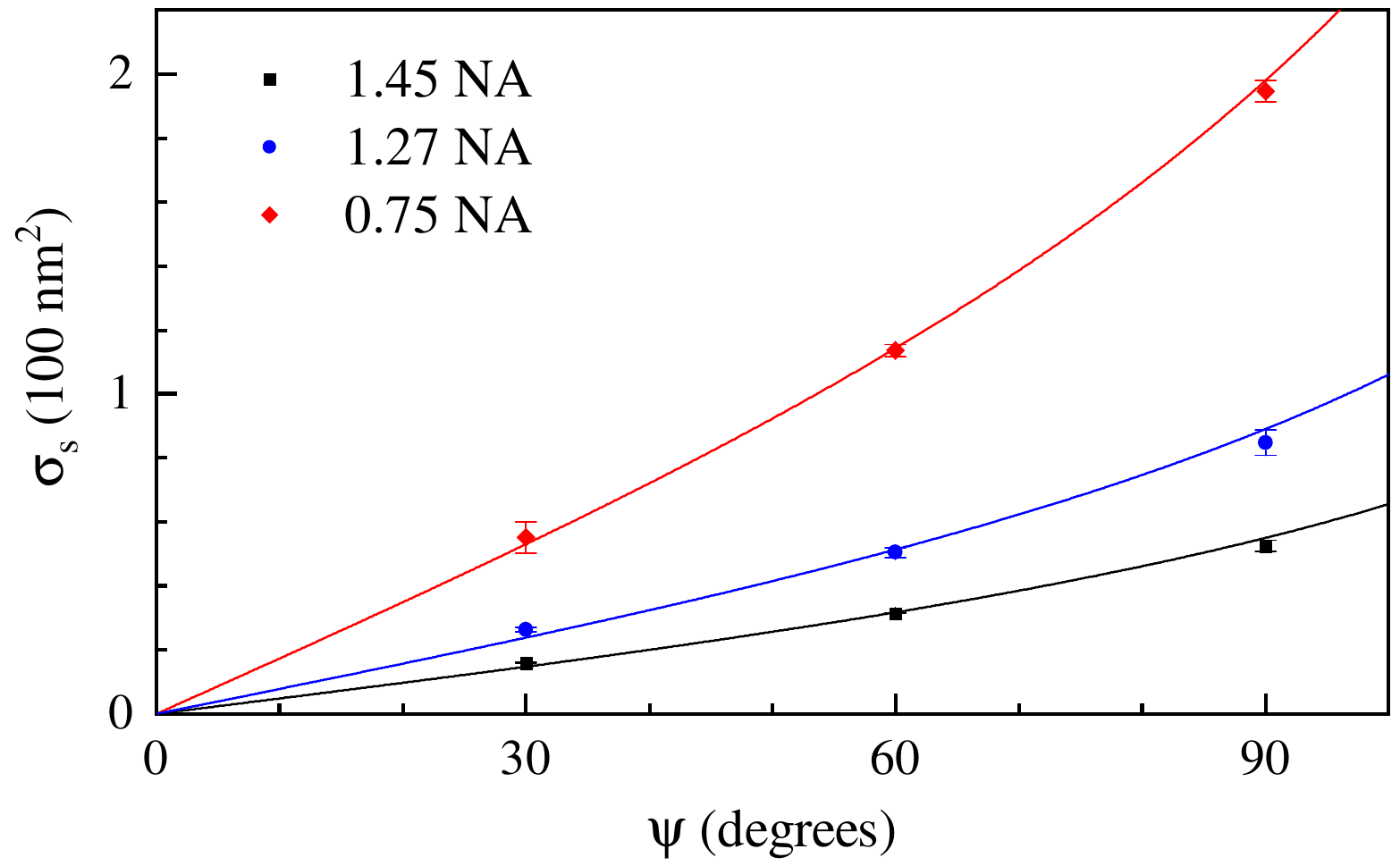}
	\caption{Single frame shot-noise \sigmasn\ of \Aphme\ for the different objectives and phase offset angles, $\psi$, when analysed using the corresponding SN pair. The lines are fits using \Eq{eq:sigmasn}.}
	\label{fig:NoisevsAngle}
\end{figure}

Recalling the scaling of the noise given in \Eq{eq:sigmad}, we fitted the dependence of the resulting \sigmasn\ for the SN pair with the phase offset $\psi$ (see \Fig{fig:NoisevsAngle}) using 
\be \label{eq:sigmasn} \sigmasn=\sigmasnz \left| \frac{1-\cos(\psi)}{\sin(\psi)}\right|\,, \ee
where $\sigmasnz$ is the noise for $\psi=90\degree$, the phase offset with the largest retrieval range in \Eq{eq:exact}. We find a good fit with $\sigmasnz=55,89,$ and $198$\,nm$^2$ for the 1.45\,NA, 1.27\,NA, and 0.75\,NA objectives, respectively.

As can be seen in \Tab{tab:ParticleSizes}, the smallest single frame shot noise \sigmasn\ is found for the 1.27\,NA objective and SN pair for beads in silicon oil, yielding about 14\,nm$^2$. Using \Eq{eq:AphV} (and taking into account the correction factor $\varrho$ to scale the phase area), this corresponds to a PS bead of 25\,nm radius. For samples in water oil, \sigmasn\ is 16\,nm$^2$ using the 1.45\,NA objective, which gives a PS bead radius of around 17.5\,nm, as size sensitivity limit from shot noise with a single frame acquisition. Generally, \sigmasn\ decreases with i) increasing NA due to the high spatial resolution, ii) going from SE to SN pair due to better SNR (\Fig{fig:kappaRiDep}), iii) decreasing phase offset due to the increased contrast (\Eq{eq:sigmad}). The size limit scales with the third root of the noise, and decreases going from silicone to water oil due to the increased refractive index difference (\Eq{eq:AphV}).

Ultimately, for a sufficient number of frames \Na, the shot-noise can be decreased to a point where the background noise \sigmab\ limits the sensitivity. For $\psi=30\degree$, the SN pair, and the 1.27\,NA objective, $\sigmab=(0.62 \pm 0.06)$\,nm$^2$ for silicone oil, corresponding to a smallest detectable bead radius of 8.8\,nm. For samples immersed in water oil instead, we find $\sigmab=(1.56 \pm 0.06)$\,nm$^2$ and a smallest detectable bead radius of 8.7\,nm. The similar radii would be expected if \sigmab\ would be caused by dielectric debris on the surface of a refractive index similar to the one of PS.

It should be noted that the iSCAT technique\cite{YoungS18} avoids the static background noise \sigmab\ by analyzing particles which attach and/or detach during measurements, so that the difference can be detected. Such a method can also be applied to qDIC, resulting in a sensitivity only limited by the shot-noise. For example, using the 1.45\,NA objective, samples in water oil, $\psi=30\degree$, the SN pair, and $\Na=1000$ acquisitions, the radius limit (scaling with $\Na^{-1/6}$) is down to 3.8\,nm, which can be achieved within 1\,s with a modern camera. We also note that by reducing the phase offset, the sensitivity can be increased, see \Eq{eq:sigmad}. Assuming ideal optics and $\eta=0$, we find that for $\psi=1\degree$ the limit for $\Na=1000$ is a radius of 1.8\,nm.

\subsection{PS bead sizes}

Samples with PS beads as described in \Sec{sec:Samples} were measured and the resulting $\Aph$ was converted into a PS volume, using the refractive index of $\np=1.59$. Note that this index can vary depending on the packing density of the PS. Hence, rather than assuming a nominal value, we have determined the refractive index for the PS beads used here, by considering the measured change of the DIC contrast versus immersion medium index, as discussed in the ESI \Sec{S-sec:PSindex}. 

The retrieved particle radii, using PS beads in silicone oil and the 1.45\,NA objective, are shown in \Fig{fig:CorrFacRadius} for the SE and SN pairs. 
The histograms were fitted with \Eq{eq:FitFunction} yielding $(R_1\pm\sigma_1)= (101.2\,\pm\,2.2)$\,nm and $(100.5 \pm 3.0)$\,nm, respectively, as mentioned previously and summarised in \Tab{tab:ParticleSizes}. Results for other objectives and immersion oils are also given in in \Tab{tab:ParticleSizes}, with the histograms shown in the ESI \Fig{S-fig:CorrFacRadius_20x_WO} to \Fig{S-fig:CorrFacRadius_60x_SO}. Importantly, we find a quantitative agreement of the measured radii for all immersion oils, objectives, and phase offset combinations within a few \%.

The smallest $\sigma_1$ is found using the 1.45\,NA objective with beads in silicone oil and for the SN pair, which gives a relative variation of $\sigma_1/R_1=2.2\%$, below the 3\% standard deviation specified by the manufacturer. This suggests that the contribution of the measurement noise to the uncertainty in the size distribution is negligible. The influence of the measurement noise $\sigma$ to the size distribution can also be calculated, and in turn removed, resulting in a corrected \sigmaoc\ given by
\be \sigmaoc = \sigma_1\sqrt{1-\left(\frac{\sigma}{\sigmaoA}\right)^2}\,, \ee
where \sigmaoA\ is the noise in \Aphme\ corresponding to $\sigma_1$. Using \Eq{eq:AphR} this is given by  
\be \sigmaoA = \sigma_1\frac{\partial\Aphme}{\partial R} = \sigma_1\frac{8 \pi^2 R^2 (\np-\nm)}{\lambda\varrho}\,, \ee
where $\varrho$ is the mean correction factor and $R$ is the mean radius $R_1$. It was found that the influence of the measurement noise was negligible, for all objectives and both the SN and SE pairs. For the case of PS beads mounted in silicone oil, imaged using the 1.45\,NA objective at $\psi=30^\circ$, $\sigma_1$ was found to be 3.0\,nm for the SE pair and 2.2\,nm for the SN pair. The calculated $\sigma_1^c$ for these cases were found to be 2.9\,nm and 2.2\,nm for the SE and SN pairs, respectively. 

\section{Nanodiamond Volume Measurement}
\label{sec:NDs} 

\begin{figure}
	\centering
	\includegraphics[width=\columnwidth]{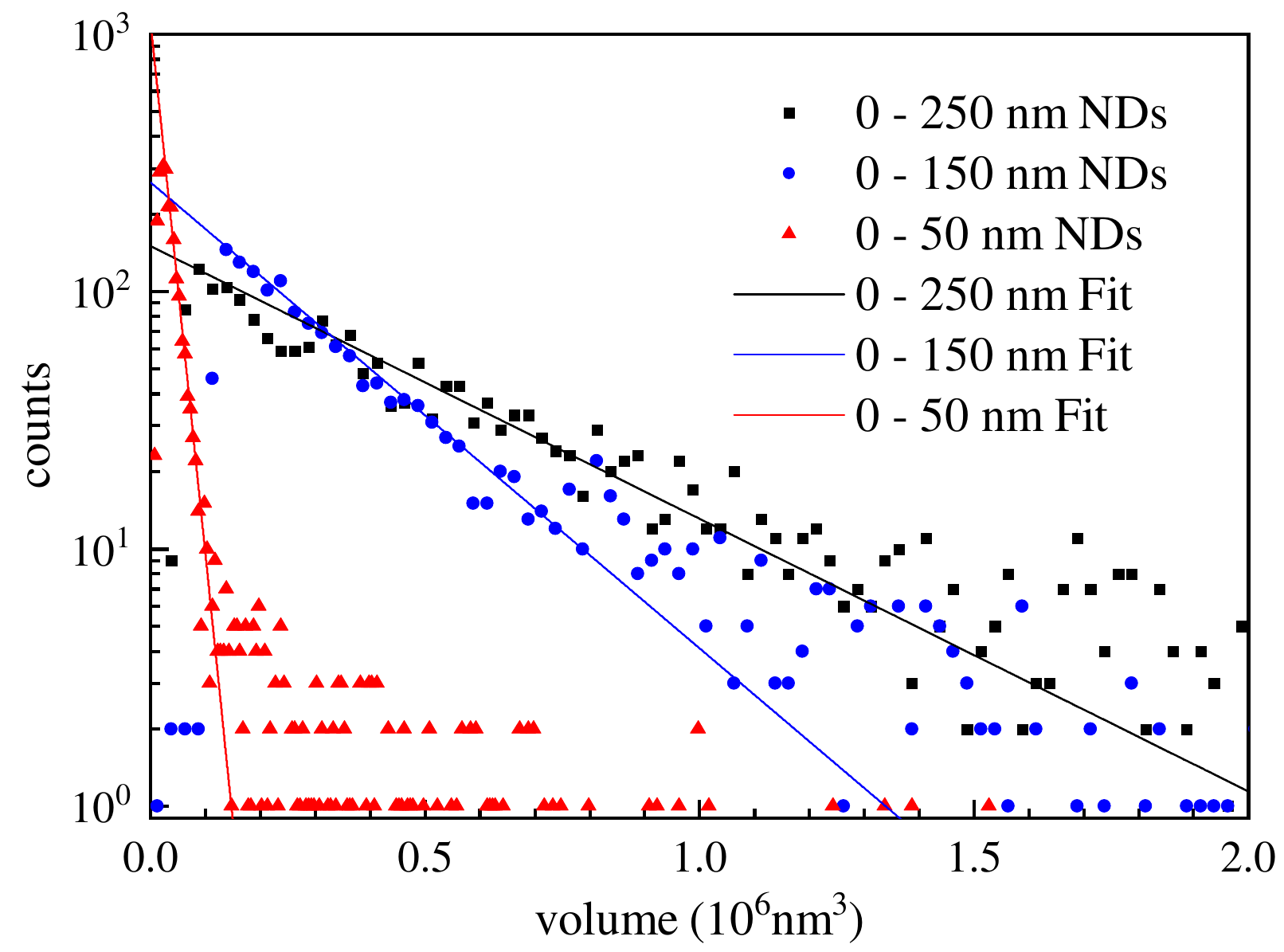}
	\caption{Size histograms measured with qDIC on nanodiamonds in silicon oil. Particles with nominal diameter ranges $(0-250)$\,nm and $(0-150)$\,nm were measured using the 0.75\,NA objective, while those with $(0-50)$\,nm diameters were imaged with the 1.27\,NA objective. The SE pair was used for the analysis. Solid lines are the exponential fits.}
	\label{fig:NDVolume}
\end{figure}

To showcase the method with an application example, we measured the size of individual nanodiamonds, specified by the manufacturer to have a broad distribution of sizes below 50\,nm, 150\,nm, and 250\,nm (see also Methods section). NDs were embedded in silicon oil. Particles with nominal diameter ranges $(0-250)$\,nm and $(0-150)$\,nm were measured using the 0.75\,NA objective, while those with $(0-50)$\,nm diameters were imaged with the 1.27\,NA objective (examples of the $\delta(\br)$ and $\phi(\br)$ images for each are shown in the ESI \Fig{S-fig:0_25NDs} to \ref{S-fig:0_05NDs}). Stacks of $\Na=256$ images were obtained at $\psi=30^\circ$ for the $(0-250)$\,nm and $(0-150)$\,nm NDs, and at $\psi=60^\circ$ for the $(0-50)$\,nm NDs. Analysis was carried out using the SE pairs.  The resulting particle volume histograms using a ND refractive index\cite{TurriOME2017} of $n=2.42$ 
show a nearly exponential decay, and thus were fitted with an exponential distribution $p_0\exp(-\Vp/\overline{V})$, with the mean volume $\overline{V}$. NDs are typically brick shaped, hence we defined a characteristic particle size using a cube geometry as $S=\sqrt[3]{\overline{V}}$. \Fig{fig:NDVolume} shows the volume distributions for each ND sample and the exponential fits. The mean volumes $\overline{V}$ were found to be $2.1 \times 10^{4}$\,nm$^3$, $2.4 \times 10^{5}$\,nm$^3$ and $4.1 \times 10^{5}$\,nm$^3$, for the $(0-50)$\,nm, $(0-150)$\,nm and $(0-250)$\,nm NDs, respectively, yielding characteristic sizes $S$ of 27.6\,nm, 62.1\,nm, and 74.3\,nm. 

\section{Conclusion}
In conclusion, we have investigated the application of quantitative DIC microscopy with Wiener filtering for sizing individual dielectric NPs, and determined the precision and accuracy of the method. Using polystyrene beads of 100\,nm radius as size standard, we found that the accuracy in determining their radius was within few nm, corresponding to a relative accuracy of only a few percent. In terms of precision, we determined the smallest detectable PS bead radius to be 7.9\,nm using a 1.27\,NA objective when mounted in silicone oil and a 1.45\,NA objective when mounted in water oil, limited by background structure at the glass interface onto which the nanoparticles were deposited. Notably, this limit can be overcome when observing particles which attach and/or detach from a glass surface during measurements, eventually reaching a sensitivity only limited by shot noise. The latter was found to equate to 3.8\,nm PS bead radius when averaging over 1000 frames, which can be achieved within 1\,s total acquisition time with modern cameras. Such sensitivity could be further increased by using small phase offsets in the DIC acquisition, potentially reaching a size limit down to only 1.8\,nm radius.
As application example, we demonstrated sizing of individual nanodiamonds having poly-disperse distributions. Small nanodiamonds with nominal sizes below 50\,nm were well above the detection limit, and were found to have a nearly exponential size distribution with 28\,nm mean size. 

Considering the importance of dielectric nanoparticles in many fields, from naturally occurring virions and exosomes to polluting nanoplastics, the proposed method could offer a powerful tool for nanoparticle analysis, combining accuracy, sensitivity and high-throughput with widely available and easy-to-use DIC microscopy.

\section*{Author contribution}

W.L. and P.B. conceived the work. S.H. prepared the samples for the optical measurements and performed the related measurements and data analysis. D.R. and L.P. supported the data analysis. S.H, P.B and W.L. wrote the manuscript. All authors contributed to the data interpretation and manuscript review.

\section*{Data availability}
Information on the data underpinning the results presented here, including how to access them, can be found in the Cardiff University data catalogue at http://doi.org/10.17035/d.2021.xxxx.

\section*{Conflicts of interest}
There are no conflicts to declare.

\section*{Acknowledgements}
S.H. acknowledges support for his PhD studies by the EPSRC Diamond Science and Technology CDT [grant n. EP/L015315/1] and Cardiff University. The microscope equipment was supported by the EPSRC grant n. EP/M028313/1. Joseph Bleddyn Williams is acknowledged for contributing to the development of the qDIC analysis software. Iestyn Pope is acknowledged for support of the microscope instrumentation.

\renewcommand{\arraystretch}{1.1}
\begin{table*}[]
	\caption{qDIC results for PS beads: correction factors $\varrho$, particle integrated phase area \Aph\ and radii $R$ with standard deviations, as well as shot noise \sigmasn\ and background noise $\sigmab$. Results for different mounting media, objective NA, phase offset $\psi$, and analysis pair are shown. The radius limit was calculated from $\sigmab$ using \Eq{eq:AphR} and $\Aph=\varrho\sigmab$. }
	\label{tab:ParticleSizes}
	\centering
\begin{tabular}{|c|c|c|c|c|c|c|c|c|c|c|}
	\hline
	$\psi$ & NA & Pair & $\kappa$ & $r_i$ & $\varrho$ & \begin{tabular}[c]{@{}c@{}}$\Aph$\\ (nm$^2$)\end{tabular} & \begin{tabular}[c]{@{}c@{}}R\\ (nm)\end{tabular} & \begin{tabular}[c]{@{}c@{}}$\sigmasn$\\ (nm$^2$)\end{tabular} & \begin{tabular}[c]{@{}c@{}}$\sigma_b$\\ (nm$^2$)\end{tabular} & \begin{tabular}[c]{@{}c@{}} Radius limit\\ (nm)\end{tabular} \\ \hline
	\multicolumn{11}{|c|}{silicone oil $\nso$=1.518} \\ \hline
	\multirow{6}{*}{30} & \multirow{2}{*}{0.75} & SN & 1 & 1 & 7.07 & 3193.4 & 97.5 $\pm$ 4.30 & 62.0 $\pm$ 1.3 & 2.05 $\pm$ 0.32 & 16 \\ \cline{3-11} 
	&  & SE & 100 & 2 & 1.27 & 3164 & 97.2 $\pm$ 4.5 & 634.7 $\pm$ 4.7 & 50.39 $\pm$ 0.85 & 27 \\ \cline{2-11} 
	& \multirow{2}{*}{1.27} & SN & 1 & 1.5 & 3.76 & 3466.1 & 100.2 $\pm$ 4.4 & 13.61 $\pm$ 0.41 & 0.62 $\pm$ 0.09 & 8.8 \\ \cline{3-11} 
	&  & SE & 100 & 2 & 1.63 & 3539.3 & 100.9 $\pm$ 3.5 & 110.7 $\pm$ 2.6 & 9.36 $\pm$ 0.44 & 16 \\ \cline{2-11} 
	& \multirow{2}{*}{1.45} & SN & 1 & 2.5 & 3.48 & 3570.9 & 101.2 $\pm$ 2.2 & 17.9 $\pm$ 1.0 & 1.94 $\pm$ 0.17 & 13 \\ \cline{3-11} 
	&  & SE & 200 & 4 & 1.25 & 3497.3 & 100.5 $\pm$ 3.0 & 131.3 $\pm$ 5.1 & 16.66 $\pm$ 0.87 & 18 \\ \hline
	\multirow{6}{*}{60} & \multirow{2}{*}{0.75} & SN & 1 & 1 & 7.07 & 4045.7 & 105.5 $\pm$ 5.7 & 121.5 $\pm$ 3.9 & 3.9 $\pm$ 1.0 & 20 \\ \cline{3-11} 
	&  & SE & 100 & 2 & 1.27 & 3518.2 & 100.7 $\pm$ 3.4 & 1250.6 $\pm$ 10.5 & 58.6 $\pm$ 2.1 & 28 \\ \cline{2-11} 
	& \multirow{2}{*}{1.27} & SN & 1 & 1.5 & 3.76 & 2907.6 & 94.5 $\pm$ 4.1 & 28.11 $\pm$ 0.55 & 0.46 $\pm$ 0.25 & 7.9 \\ \cline{3-11} 
	&  & SE & 100 & 2 & 1.63 & 3105.8 & 96.6 $\pm$ 4.2 & 235.3 $\pm$ 6.5 & 10.1 $\pm$ 1.4 & 17 \\ \cline{2-11} 
	& \multirow{2}{*}{1.45} & SN & 1 & 2.5 & 3.48 & 3549.8 & 101.0 $\pm$ 2.3 & 31.21 $\pm$ 0.50 & 1.53 $\pm$ 0.10 & 12 \\ \cline{3-11} 
	&  & SE & 200 & 4 & 1.25 & 3486.9 & 100.4 $\pm$ 2.5 & 215.1 $\pm$ 8.8 & 18.4 $\pm$ 1.5 & 19 \\ \hline
	\multirow{6}{*}{90} & \multirow{2}{*}{0.75} & SN & 1 & 1 & 7.07 & 4011.3 & 105.2 $\pm$ 5.4 & 212 $\pm$ 13 & 0.0 $\pm$ 5.4 & 22 \\ \cline{3-11} 
	&  & SE & 100 & 2 & 1.27 & 3507.8 & 100.6 $\pm$ 4.4 & 2222 $\pm$ 30 & 77.2 $\pm$ 7.3 & 31 \\ \cline{2-11} 
	& \multirow{2}{*}{1.27} & SN & 1 & 1.5 & 3.76 & 2639 & 91.5 $\pm$ 4.2 & 49.17 $\pm$ 0.55 & 1.17 $\pm$ 0.18 & 11 \\ \cline{3-11} 
	&  & SE & 100 & 2 & 1.63 & 2735.7 & 92.6 $\pm$ 4.7 & 375.4 $\pm$ 3.9 & 14.14 $\pm$ 0.91 & 19 \\ \cline{2-11} 
	& \multirow{2}{*}{1.45} & SN & 1 & 2.5 & 3.48 & 3645.5 & 101.9 $\pm$ 2.1 & 51.1 $\pm$ 1.8 & 2.27 $\pm$ 0.38 & 13 \\ \cline{3-11} 
	&  & SE & 200 & 4 & 1.25 & 3602.8 & 101.5 $\pm$ 2.5 & 361 $\pm$ 13 & 24 $\pm$ 24 & 21 \\ \hline
	\multicolumn{11}{|c|}{water immersion oil $\nwo$=1.334} \\ \hline
	\multirow{6}{*}{30} & \multirow{2}{*}{0.75} & SN & 1 & 1 & 6.75 & 13000.1 & 102.0 $\pm$ 4.3 & 61.8 $\pm$ 1.4 & 4.00 $\pm$ 0.24 & 13 \\ \cline{3-11} 
	&  & SE & 100 & 2 & 1.23 & 12434.9 & 100.5 $\pm$ 2.5 & 644 $\pm$ 17 & 91.9 $\pm$ 3.0 & 21 \\ \cline{2-11} 
	& \multirow{2}{*}{1.27} & SN & 1 & 1.5 & 5.26 & 12103.9 & 99.6 $\pm$ 4.2 & 24.03 $\pm$ 0.54 & 1.56 $\pm$ 0.09 & 8.7 \\ \cline{3-11} 
	&  & SE & 100 & 2 & 2.12 & 11922.5 & 99.1 $\pm$ 3.7 & 106.94 $\pm$ 0.86 & 10.52 $\pm$ 0.14 & 12 \\ \cline{2-11} 
	& \multirow{2}{*}{1.45} & SN & 1 & 2.5 & 4.12 & 12472.1 & 100.6 $\pm$ 5.5 & 15.98 $\pm$ 0.37 & 1.45 $\pm$ 0.06 & 7.9 \\ \cline{3-11} 
	&  & SE & 200 & 4 & 1.38 & 12711.7 & 101.2 $\pm$ 4.3 & 139 $\pm$ 45 & 17.7 $\pm$ 7.2 & 13 \\ \hline
	\multirow{6}{*}{60} & \multirow{2}{*}{0.75} & SN & 1 & 1 & 6.75 & 12031.1 & 99.4 $\pm$ 3.6 & 122.3 $\pm$ 3.8 & 4.66 $\pm$ 0.88 & 14 \\ \cline{3-11} 
	&  & SE & 100 & 2 & 1.23 & 10940.2 & 96.3 $\pm$ 2.8 & 1276 $\pm$ 14 & 100.0 $\pm$ 2.4 & 22 \\ \cline{2-11} 
	& \multirow{2}{*}{1.27} & SN & 1 & 1.5 & 5.26 & 10838.3 & 96.0 $\pm$ 5.1 & 50.6 $\pm$ 1.4 & 2.07 $\pm$ 0.29 & 9.6 \\ \cline{3-11} 
	&  & SE & 100 & 2 & 2.12 & 11008.5 & 96.5 $\pm$ 4.5 & 227.7 $\pm$ 3.4 & 11.83 $\pm$ 0.65 & 13 \\ \cline{2-11} 
	& \multirow{2}{*}{1.45} & SN & 1 & 2.5 & 4.12 & 14019.8 & 104.6 $\pm$ 5.6 & 32.43 $\pm$ 0.80 & 1.56 $\pm$ 0.16 & 8.1 \\ \cline{3-11} 
	&  & SE & 200 & 4 & 1.38 & 15431.8 & 108.0 $\pm$ 4.2 & 233 $\pm$ 12 & 18.7 $\pm$ 2.0 & 13 \\ \hline
	\multirow{6}{*}{90} & \multirow{2}{*}{0.75} & SN & 1 & 1 & 6.75 & 11145.9 & 96.9 $\pm$ 4.2 & 197.2 $\pm$ 6.5 & 3.1 $\pm$ 3.1 & 12 \\ \cline{3-11} 
	&  & SE & 100 & 2 & 1.23 & 10272.6 & 94.3 $\pm$ 2.9 & 2108 $\pm$ 61 & 141 $\pm$ 11 & 24 \\ \cline{2-11} 
	& \multirow{2}{*}{1.27} & SN & 1 & 1.5 & 5.26 & 11249.8 & 97.2 $\pm$ 5.4 & 86.8 $\pm$ 3.6 & 2.5 $\pm$ 1.1 & 10 \\ \cline{3-11} 
	&  & SE & 100 & 2 & 2.12 & 11284.6 & 97.3 $\pm$ 4.6 & 395 $\pm$ 13 & 11.8 $\pm$ 3.5 & 13 \\ \cline{2-11} 
	& \multirow{2}{*}{1.45} & SN & 1 & 2.5 & 4.12 & 14923.1 & 106.8 $\pm$ 4.4 & 53.8 $\pm$ 1.7 & 2.39 $\pm$ 0.35 & 9.3 \\ \cline{3-11} 
	&  & SE & 200 & 4 & 1.38 & 14425.7 & 105.6 $\pm$ 5.7 & 370 $\pm$ 13 & 27.2 $\pm$ 2.2 & 15 \\ \hline
\end{tabular}%

\end{table*}

\providecommand*{\mcitethebibliography}{\thebibliography}
\csname @ifundefined\endcsname{endmcitethebibliography}
{\let\endmcitethebibliography\endthebibliography}{}


\begin{mcitethebibliography}{25}
	\providecommand*{\natexlab}[1]{#1}
	\providecommand*{\mciteSetBstSublistMode}[1]{}
	\providecommand*{\mciteSetBstMaxWidthForm}[2]{}
	\providecommand*{\mciteBstWouldAddEndPuncttrue}
	{\def\EndOfBibitem{\unskip.}}
	\providecommand*{\mciteBstWouldAddEndPunctfalse}
	{\let\EndOfBibitem\relax}
	\providecommand*{\mciteSetBstMidEndSepPunct}[3]{}
	\providecommand*{\mciteSetBstSublistLabelBeginEnd}[3]{}
	\providecommand*{\EndOfBibitem}{}
	\mciteSetBstSublistMode{f}
	\mciteSetBstMaxWidthForm{subitem}
	{(\emph{\alph{mcitesubitemcount}})}
	\mciteSetBstSublistLabelBeginEnd{\mcitemaxwidthsubitemform\space}
	{\relax}{\relax}
	
	\bibitem[Sun \emph{et~al.}(2014)Sun, Zhang, Pang, Hyun, Yang, and
	Xia]{SunACIE14}
	T.~Sun, Y.~S. Zhang, B.~Pang, D.~C. Hyun, M.~Yang and Y.~Xia, \emph{Angew.
		Chem. Int. Ed.}, 2014, \textbf{53}, 12320 --12364\relax
	\mciteBstWouldAddEndPuncttrue
	\mciteSetBstMidEndSepPunct{\mcitedefaultmidpunct}
	{\mcitedefaultendpunct}{\mcitedefaultseppunct}\relax
	\EndOfBibitem
	\bibitem[Liu \emph{et~al.}(2019)Liu, Zhang, and Xu]{LiuSmall19}
	J.~Liu, R.~Zhang and Z.~P. Xu, \emph{Small}, 2019, \textbf{15}, 1900262\relax
	\mciteBstWouldAddEndPuncttrue
	\mciteSetBstMidEndSepPunct{\mcitedefaultmidpunct}
	{\mcitedefaultendpunct}{\mcitedefaultseppunct}\relax
	\EndOfBibitem
	\bibitem[Dang \emph{et~al.}(2013)Dang, Yuan, Yao, and Liao]{DangAM13}
	Z.-M. Dang, J.-K. Yuan, S.-H. Yao and R.-J. Liao, \emph{Advanced Materials},
	2013, \textbf{25}, 6334--6365\relax
	\mciteBstWouldAddEndPuncttrue
	\mciteSetBstMidEndSepPunct{\mcitedefaultmidpunct}
	{\mcitedefaultendpunct}{\mcitedefaultseppunct}\relax
	\EndOfBibitem
	\bibitem[Mourdikoudis \emph{et~al.}(2018)Mourdikoudis, Pallares, and
	Thanh]{MourdikoudisNanoscale18}
	S.~Mourdikoudis, R.~M. Pallares and N.~T.~K. Thanh, \emph{Nanoscale}, 2018,
	\textbf{10}, 12871--12934\relax
	\mciteBstWouldAddEndPuncttrue
	\mciteSetBstMidEndSepPunct{\mcitedefaultmidpunct}
	{\mcitedefaultendpunct}{\mcitedefaultseppunct}\relax
	\EndOfBibitem
	\bibitem[Payne \emph{et~al.}(2020)Payne, Albrecht, Langbein, and
	Borri]{PayneNS20}
	L.~M. Payne, W.~Albrecht, W.~Langbein and P.~Borri, \emph{Nanoscale}, 2020,
	\textbf{12}, 16215--16228\relax
	\mciteBstWouldAddEndPuncttrue
	\mciteSetBstMidEndSepPunct{\mcitedefaultmidpunct}
	{\mcitedefaultendpunct}{\mcitedefaultseppunct}\relax
	\EndOfBibitem
	\bibitem[Wang \emph{et~al.}(2020)Wang, Zilli, Sztranyovszky, Langbein, and
	Borri]{WangNSA20}
	Y.~Wang, A.~Zilli, Z.~Sztranyovszky, W.~Langbein and P.~Borri, \emph{Nanoscale
		Adv.}, 2020, \textbf{2}, 2485--2496\relax
	\mciteBstWouldAddEndPuncttrue
	\mciteSetBstMidEndSepPunct{\mcitedefaultmidpunct}
	{\mcitedefaultendpunct}{\mcitedefaultseppunct}\relax
	\EndOfBibitem
	\bibitem[Payne \emph{et~al.}(2018)Payne, Langbein, and Borri]{PaynePRAP18}
	L.~M. Payne, W.~Langbein and P.~Borri, \emph{Phys. Rev. Appl}, 2018,
	\textbf{9}, 034006\relax
	\mciteBstWouldAddEndPuncttrue
	\mciteSetBstMidEndSepPunct{\mcitedefaultmidpunct}
	{\mcitedefaultendpunct}{\mcitedefaultseppunct}\relax
	\EndOfBibitem
	\bibitem[Young \emph{et~al.}(2018)Young, Hundt, Cole, Fineberg, Andrecka,
	Tyler, Olerinyova, Ansari, Marklund, Collier, Chandler, Tkachenko, Allen,
	Crispin, Billington, Takagi, Sellers, Eichmann, Selenko, Frey, Riek, Galpin,
	Struwe, Benesch, and Kukura]{YoungS18}
	G.~Young, N.~Hundt, D.~Cole, A.~Fineberg, J.~Andrecka, A.~Tyler, A.~Olerinyova,
	A.~Ansari, E.~G. Marklund, M.~P. Collier, S.~A. Chandler, O.~Tkachenko,
	J.~Allen, M.~Crispin, N.~Billington, Y.~Takagi, J.~R. Sellers, C.~Eichmann,
	P.~Selenko, L.~Frey, R.~Riek, M.~R. Galpin, W.~B. Struwe, J.~L.~P. Benesch
	and P.~Kukura, \emph{Science}, 2018, \textbf{360}, 423--427\relax
	\mciteBstWouldAddEndPuncttrue
	\mciteSetBstMidEndSepPunct{\mcitedefaultmidpunct}
	{\mcitedefaultendpunct}{\mcitedefaultseppunct}\relax
	\EndOfBibitem
	\bibitem[Taylor and Sandoghdar(2019)]{TaylorNL19}
	R.~W. Taylor and V.~Sandoghdar, \emph{Nano Lett.}, 2019, \textbf{19},
	4827--4835\relax
	\mciteBstWouldAddEndPuncttrue
	\mciteSetBstMidEndSepPunct{\mcitedefaultmidpunct}
	{\mcitedefaultendpunct}{\mcitedefaultseppunct}\relax
	\EndOfBibitem
	\bibitem[Nomarski(1955)]{NomarskiJPRP55}
	G.~M. Nomarski, \emph{J. Phys. Radium Paris}, 1955, \textbf{16}, 9S\relax
	\mciteBstWouldAddEndPuncttrue
	\mciteSetBstMidEndSepPunct{\mcitedefaultmidpunct}
	{\mcitedefaultendpunct}{\mcitedefaultseppunct}\relax
	\EndOfBibitem
	\bibitem[Arnison \emph{et~al.}(2004)Arnison, Larkin, Sheppard, Smith, and
	Cogswell]{ArnisonJM04}
	M.~R. Arnison, K.~G. Larkin, C.~J.~R. Sheppard, N.~I. Smith and C.~J. Cogswell,
	\emph{J. Microsc.}, 2004, \textbf{214}, 7--12\relax
	\mciteBstWouldAddEndPuncttrue
	\mciteSetBstMidEndSepPunct{\mcitedefaultmidpunct}
	{\mcitedefaultendpunct}{\mcitedefaultseppunct}\relax
	\EndOfBibitem
	\bibitem[King \emph{et~al.}(2008)King, Libertun, Piestun, Cogswell, and
	Preza]{KingJBO08}
	S.~V. King, A.~Libertun, R.~Piestun, C.~J. Cogswell and C.~Preza, \emph{J.
		Biomed. Opt.}, 2008, \textbf{13}, 024020\relax
	\mciteBstWouldAddEndPuncttrue
	\mciteSetBstMidEndSepPunct{\mcitedefaultmidpunct}
	{\mcitedefaultendpunct}{\mcitedefaultseppunct}\relax
	\EndOfBibitem
	\bibitem[Duncan \emph{et~al.}(2011)Duncan, Fischer, Dayton, and
	Prahl]{DuncanJOSAA11}
	D.~D. Duncan, D.~G. Fischer, A.~Dayton and S.~A. Prahl, \emph{Journal of the
		Optical Society of America A}, 2011, \textbf{28}, 1297\relax
	\mciteBstWouldAddEndPuncttrue
	\mciteSetBstMidEndSepPunct{\mcitedefaultmidpunct}
	{\mcitedefaultendpunct}{\mcitedefaultseppunct}\relax
	\EndOfBibitem
	\bibitem[Kou and Sheppard(2010)]{KouACP10}
	S.~S. Kou and C.~Sheppard, International Conference on Advanced Phase
	Measurements Methods in Optics and Imaging, 2010, pp. 301--306\relax
	\mciteBstWouldAddEndPuncttrue
	\mciteSetBstMidEndSepPunct{\mcitedefaultmidpunct}
	{\mcitedefaultendpunct}{\mcitedefaultseppunct}\relax
	\EndOfBibitem
	\bibitem[Shribak \emph{et~al.}(2017)Shribak, Larkin, and Biggs]{ShribakJBO17}
	M.~Shribak, K.~G. Larkin and D.~Biggs, \emph{J. Biomed. Opt.}, 2017,
	\textbf{22}, 016006\relax
	\mciteBstWouldAddEndPuncttrue
	\mciteSetBstMidEndSepPunct{\mcitedefaultmidpunct}
	{\mcitedefaultendpunct}{\mcitedefaultseppunct}\relax
	\EndOfBibitem
	\bibitem[Ding \emph{et~al.}(2019)Ding, Li, Deng, and Simpson]{DingOE19}
	C.~Ding, C.~Li, F.~Deng and G.~J. Simpson, \emph{Optics Express}, 2019,
	\textbf{27}, 3837\relax
	\mciteBstWouldAddEndPuncttrue
	\mciteSetBstMidEndSepPunct{\mcitedefaultmidpunct}
	{\mcitedefaultendpunct}{\mcitedefaultseppunct}\relax
	\EndOfBibitem
	\bibitem[van Munster \emph{et~al.}(1997)van Munster, van Vliet, and
	Aten]{MunsterJM97}
	E.~B. van Munster, L.~J. van Vliet and J.~A. Aten, \emph{J. Microsc.}, 1997,
	\textbf{188}, 149--157\relax
	\mciteBstWouldAddEndPuncttrue
	\mciteSetBstMidEndSepPunct{\mcitedefaultmidpunct}
	{\mcitedefaultendpunct}{\mcitedefaultseppunct}\relax
	\EndOfBibitem
	\bibitem[Koos \emph{et~al.}(2016)Koos, Moln\'{a}r, Kelemen, Tam\'{a}s, and
	Horvath]{KoosSR16}
	K.~Koos, J.~Moln\'{a}r, L.~Kelemen, G.~Tam\'{a}s and P.~Horvath, \emph{Sci.
		Rep.}, 2016, \textbf{6}, 30420\relax
	\mciteBstWouldAddEndPuncttrue
	\mciteSetBstMidEndSepPunct{\mcitedefaultmidpunct}
	{\mcitedefaultendpunct}{\mcitedefaultseppunct}\relax
	\EndOfBibitem
	\bibitem[Regan \emph{et~al.}(2019)Regan, Williams, Borri, and
	Langbein]{ReganL19}
	D.~Regan, J.~Williams, P.~Borri and W.~Langbein, \emph{Langmuir}, 2019,
	\textbf{35}, 13805--13814\relax
	\mciteBstWouldAddEndPuncttrue
	\mciteSetBstMidEndSepPunct{\mcitedefaultmidpunct}
	{\mcitedefaultendpunct}{\mcitedefaultseppunct}\relax
	\EndOfBibitem
	\bibitem[Regan \emph{et~al.}(2019)Regan, Williams, Masia, Borri, and
	Langbein]{ReganSPIE19}
	D.~Regan, J.~Williams, F.~Masia, P.~Borri and W.~Langbein, Quantitative Phase
	Imaging V, 2019\relax
	\mciteBstWouldAddEndPuncttrue
	\mciteSetBstMidEndSepPunct{\mcitedefaultmidpunct}
	{\mcitedefaultendpunct}{\mcitedefaultseppunct}\relax
	\EndOfBibitem
	\bibitem[McPhee \emph{et~al.}(2013)McPhee, Zoriniants, Langbein, and
	Borri]{McPheeBJ13}
	C.~I. McPhee, G.~Zoriniants, W.~Langbein and P.~Borri, \emph{Biophys. J.},
	2013, \textbf{105}, 1414--1420\relax
	\mciteBstWouldAddEndPuncttrue
	\mciteSetBstMidEndSepPunct{\mcitedefaultmidpunct}
	{\mcitedefaultendpunct}{\mcitedefaultseppunct}\relax
	\EndOfBibitem
	\bibitem[Pope \emph{et~al.}(2014)Pope, Payne, Zoriniants, Thomas, Williams,
	Watson, Langbein, and Borri]{PopeNNa14}
	I.~Pope, L.~Payne, G.~Zoriniants, E.~Thomas, O.~Williams, P.~Watson,
	W.~Langbein and P.~Borri, \emph{Nat. Nanotechnol.}, 2014, \textbf{9},
	940--946\relax
	\mciteBstWouldAddEndPuncttrue
	\mciteSetBstMidEndSepPunct{\mcitedefaultmidpunct}
	{\mcitedefaultendpunct}{\mcitedefaultseppunct}\relax
	\EndOfBibitem
	\bibitem[Payne \emph{et~al.}(2013)Payne, Langbein, and Borri]{PayneAPL13}
	L.~M. Payne, W.~Langbein and P.~Borri, \emph{Appl. Phys. Lett.}, 2013,
	\textbf{102}, 131107\relax
	\mciteBstWouldAddEndPuncttrue
	\mciteSetBstMidEndSepPunct{\mcitedefaultmidpunct}
	{\mcitedefaultendpunct}{\mcitedefaultseppunct}\relax
	\EndOfBibitem
	\bibitem[Payne \emph{et~al.}(2015)Payne, Zoriniants, Masia, Arkill, Verkade,
	Rowles, Langbein, and Borri]{PayneFD15}
	L.~Payne, G.~Zoriniants, F.~Masia, K.~P. Arkill, P.~Verkade, D.~Rowles,
	W.~Langbein and P.~Borri, \emph{Faraday Discuss.}, 2015, \textbf{184},
	305--320\relax
	\mciteBstWouldAddEndPuncttrue
	\mciteSetBstMidEndSepPunct{\mcitedefaultmidpunct}
	{\mcitedefaultendpunct}{\mcitedefaultseppunct}\relax
	\EndOfBibitem
	\bibitem[Turri \emph{et~al.}(2017)Turri, Webster, Chen, Wickham, Bennett, and
	Bass]{TurriOME2017}
	G.~Turri, S.~Webster, Y.~Chen, B.~Wickham, A.~Bennett and M.~Bass,
	\emph{Optical Materials Express}, 2017, \textbf{7}, 855--859\relax
	\mciteBstWouldAddEndPuncttrue
	\mciteSetBstMidEndSepPunct{\mcitedefaultmidpunct}
	{\mcitedefaultendpunct}{\mcitedefaultseppunct}\relax
	\EndOfBibitem
\end{mcitethebibliography}
\end{document}